\documentclass[pre,preprint,aps,]{revtex4}
\usepackage{graphicx}
\usepackage{comment}
\usepackage{amsmath,amssymb,amsthm} 
\usepackage{bm}
\usepackage{verbatim}
\usepackage{float}

\begin{document}

		\title{ Nucleation and  droplet growth from  supersaturated vapor at temperatures below the triple point temperature}

\author{S\o ren Toxvaerd }
\affiliation{DNRF centre  ``Glass and Time,'' IMFUFA, Department
 of Sciences, Roskilde University, Postbox 260, DK-4000 Roskilde, Denmark}
\date{\today}

\vspace*{0.7cm}

\begin{abstract}
In 1897 Ostwald formulated his step rule for formation of the most stable crystal state for a system  with crystal polymorphism.
The rule describes  the irreversible way a  system converts to the crystal  with lowest free energy. But in fact the
irreversible way a supercooled gas below the triple point temperature  $T_{tr.p.}$ crystallizes via a liquid droplet is an example of Ostwald's step rule.
  The homogeneous nucleation in the supersaturated gas is not to a crystal, but to a liquid-like critical nucleus.
We have for the first time performed constant energy (NVE) Molecular Dynamics (MD) of homogeneous nucleation  without the use of a thermostat.
The simulations of homogeneous nucleation in a Lennard-Jones system from supersaturated vapor at temperatures below 
 $T_{tr.p.}$ reveals  that the nucleation to a liquid-like critical nucleus is initiated by a small cold  cluster
 [S. Toxvaerd, J. Chem. Phys. \textbf{143} 154705 (2015)].
 The release of latent heat at the subsequent droplet growth  increases the temperature
in the liquid-like droplet, which for not deep supercooling and/or low supersaturation,
 can exceed  $T_{tr.p.}$. The temperature of the liquid-like droplet increases less
for a low supersaturation and
 remains below  $T_{tr.p.}$, but without a crystallization of the droplet for long times.  The dissipation of
the latent heat into the surrounding gas is affected by a traditional MD thermostat, with the consequence 
that droplet growth is different for (NVE) MD  and constant temperature (NVT) MD. 
\end{abstract}

\maketitle

\section{Introduction}

In 1897 Ostwald formulated his step rule for  the irreversible reaction 
of a system with crystal  polymorphism  to   the most stable crystal form \cite{Ostwald}.
 But in fact the
irreversible way a supercooled gas below the triple point temperature crystallizes is an example of Ostwald's step rule.
The homogeneous critical nucleus is not a crystal, but a liquid-like droplet.

Condensation of supercooled vapor  at  temperatures below the triple point temperature
 $T_{tr.p.}$ is of general interest. E.g.
the condensation of water molecules at a very low temperature in  interstellar space might appear as
 amorphous solid form of water, consisting of water molecules that are randomly
 arranged like the molecules in common glass. \cite{Debenedetti,Angell}. But in general
one shall, however, expect that the condensation appears either by creation of a critical crystal nucleus
or, if the condensation appears by a creation of a liquid-like critical nucleus, that
the transition  shortly after is followed by a crystallization in the nucleus.

Nucleation and droplet growth has been simulated by various methods during several decades \cite{Daan,Matsumoto,Tox1} .
Realistic computer simulations of nucleation require very big systems of many particles and simulated over long times. But
  the actual supersaturations are  even
for these systems typically significant higher than what appears in nature \cite{Schweizer}.
Most simulations are performed by Molecular Dynamics (MD) simulations, and  for  systems
of Lennard-Jones (LJ) particles and with comparison of nucleation rates with corresponding rates obtained for supersaturated
Argon gas \cite{Diemand}. For nucleation of vapor particles below  $T_{tr.p.}$ one apparently observes a condensation
to a liquid-like critical nucleus  and a liquid droplet at the  subsequent droplet growth \cite{Diemand,Tanaka1} . This result is  surprising and
this is one of the motivations for the present investigation.

In a recent article it was found that the creation of a  critical nucleus  appears by the growth of a  small cold sub-critical cluster \cite{toxcold}.
This mechanism might not be so surprising, although it contradicts traditional theories and
simulations of homogeneous nucleation, which assume that the  temperature of the  supersaturated vapor ensures an isothermal
homogeneous nucleation. For this reason almost all MD simulations are performed with a thermostat. A second motivation for the present 
work is to investigate whether homogeneous nucleation  below the triple point temperature also is initiated by a small cold  nucleus.

Nucleation and droplet growth has been simulated with thermostats that ensure an overall constant temperature during
the phase change. But for the first time homogeneous nucleation and the subsequent
droplet growth is simulated without a thermostat.  The droplet growth is associated with a significant release of
latent heat, which dissipates slowly into the surrounding gas. The present simulations  with and without a thermostat
 show that the traditional MD with thermostats influence the
droplet growth, and in a nonphysical way.

Most of the present simulations are performed for a temperature $T=0.50$  below
 the triple point temperature $T_{tr.p.}=0.674$ of a LJ system \cite{Mastny}.
MD for nucleation is described in Section II, and the homogeneous nucleation at  $T=0.50$ is given in  Section III.
The succeeding droplet grow is investigated in Section IV, and the effects of different thermostats are given in the section.
The results are summarized and discussed in the last section,  Section V.

\section{ Molecular Dynamics  simulation. }

The system consists of $N=40000$ Lennard Jones (LJ)
particles in a cubic box with periodical boundaries \cite{ToxMD}. The MD simulations
are performed with the central difference algorithm in the leap-frog
 version, and the forces for pair interactions  greater than $r_{cut}$ are ignored. There are different ways
to take the non-analyticity of the potential at $r_{cut}$  into account. The most stable and energy conserving way
is to cut and shift the forces (SF) \cite{Tox2}, but most simulations are
 for cut and shifted potentials (SP). The extensive MD simulations
of homogeneous nucleation \cite{Diemand,Angelil} were performed for a SP cut with $r_{cut}= 5 \sigma$. The
  present simulations are for SF with  $r_{cut}= 5 \sigma$, and the SF ensures energy conservation even for billion of
time steps \cite{Tox3,Tox4} and  makes it possible to perform long  simulations (NVE) without a thermostat.

 The time averages of the NVE simulations corresponds to  microcanonical averages with
a constant energy. The dynamics can, however, be constrained to a  (mean)
 temperature (NVT), e.g by a Nos\'{e}-Hoover constraint (NH) \cite{Tox6}. Another way is simply to rescale the
velocities to a given mean value \cite{Diemand}. The MD are performed with a time increment $dt=0.01$
and the NH- MD with a "friction constant" $ \eta=0.05$. The sensitivity of  homogeneous nucleation with respect to
the values of  $dt$ and $ \eta$ was investigated in  \cite{toxcold}, where also  further MD technical details are given.

It is necessary to determine the distribution of  clusters and their temperatures   every
time step in order to obtain  the dynamics of nucleation. A cluster consists of particles with at least
two nearest neighbors (a "Stillinger criterion").
 The determination of clusters can be done without a significant increase in computer time \cite{toxcold}. 

The thermostats are "intensive", by adjusting the kinetic energy of a particle to the mean kinetic energy proportional
to its excess kinetic energy and independent of its position.  For the big systems  with many  particles and with a few small nuclei the difference between
the mean temperature of $all$ the particles with and without a rescaling are tiny and a thermostat has no effect on the  nucleation.
But the intensive thermostats  differ from the way a real system is affected by a local excess of kinetic energy. The
succeeding droplet growth is associated with a significant release of latent heat and the droplets have a significant higher
temperature than the surrounding supercooled gas. The dissipation of the heat into the surrounding is affected
by an intensive thermostat, and with the consequence that NVE and NVT  simulations of droplet growth differ.

\section{Homogeneous nucleation below the triple point temperature}

Homogeneous nucleation by MD is a stochastic  process,
 and it is not possible to predict when the nucleation takes place. 
The supersaturated gas can remain in the quasi-equilibrium state (QES) for very long time,
but sooner or later a critical nucleus appears and the system  phase separates. The degree of supersaturation
 $S$  at the temperature $T$ is given by the pressure $P$ or density $\rho$ in the QES state, $S=P/P_{eq} \approx \rho/ \rho_{eq}$,
where $P_{eq}$ and $\rho_{eq}$ are the pressure and gas density of coexisting (supercooled) liquid and gas at the same temperature.

 Most of the particles in the QES are  free and not bound to any other particles, but there are some dimers and  an exponential
declining distribution of  bigger clusters  \cite{toxcold}. A nucleation might appear when  a small  cluster grows and
reaches the critical nucleation size. Occasionally this event does not result in a nucleation, but most often
the growth continues. The dynamics of homogeneous nucleation can be separated into three time intervals: (I) the time where the system remains
in the QES with $\textit{sub-critical}$ $\textit{ clusters}$, (II) the time  interval where a sub-critical cluster grows to
$\textit{ a}$  $\textit{critical}$  $\textit{nucleus}$, and (III)
the state  after a stable  $\textit{droplet}$ is created and where it  grows   until
the system  reaches the equilibrium state of corresponding liquid and gas.

 The temperature of a successful sub-critical cluster fluctuates, and 
 it is necessary to perform independent 
nucleations in order to determine the mean behavior
of a nucleating  cluster. 
 As in \cite{toxcold} we have performed 25 independent homogeneous nucleations. The simulations are
  at the temperature $T=0.50$ and $\rho=0.0040$,  and the data are
collected in Table I. This temperature is well below the triple point temperature $T_{tr.p.}=0.674$ of a LJ system,
and  the corresponding condensed state is the (fcc) crystal state.
 The QES with $\rho=0.0040$ is a supersaturated state with $S \approx 73$,
and is one of the  states also investigated by \cite{Diemand,Angelil} (T5n40). The present
data for nucleation agrees with theirs, also  with respect to the size of the critical nucleus: $n_{cl} \approx 18$.
$ $\\

{\bf Table 1}. Data for the homogeneous nucleation at $T=0.50$  and QES density $\rho=0.0040$.
$ $\\
 \begin{tabbing}
 \hspace{2.9 cm}\=\hspace{2. cm}\=\hspace{2. cm}\=\hspace{2. cm}\kill
 $t_{nucl. start}$ \> $ \Delta t_{nucl}$  \> $ T_{cl}$ \> $ n_{coor.no.}$ \\
 \end{tabbing}
 \begin{tabbing}
  \hspace{2.9 cm}\=\hspace{2. cm}\=\hspace{2. cm}\=\hspace{2 cm}\kill
 67000$_{\pm 43000}$ \> 1500$_{\pm 1000 }$ \> 0.486$_{\pm 8}$ \> 7.1$_{\pm 4}$ \\
 onset    \> 200 \> 0.467$_{\pm 19}$ \> 6.1$_{\pm 3}$ \\
 \end{tabbing}
$ $\\
$t_{nucl. start}$: mean time before a successful sub-critical cluster  appears.\\
 $ \Delta t_{nucl}$:  nucleation time for  this sub-critical cluster.\\
 $ T_{cl}$:   mean temperature of the
sub-critical cluster in  $ \Delta t_{nucl}$; and at onset of nucleation. \\
$ n_{coor.no.}$: the mean coordination number of a particle at the center of the sub-critical cluster.\\

\begin{figure}
\begin{center}
\includegraphics[width=12cm,angle=0]{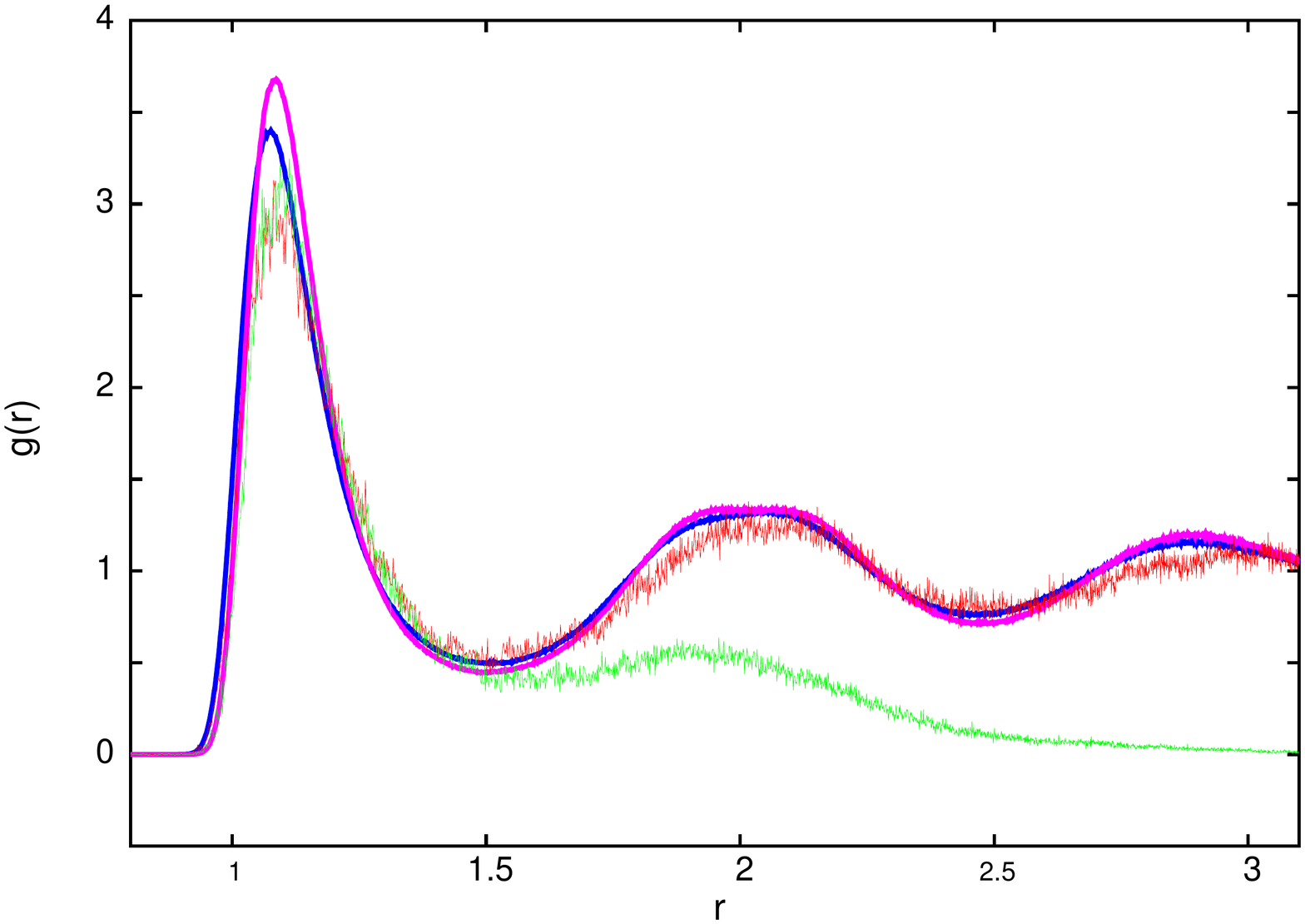}

\end{center}
\caption{Radial distribution $g(r)$ of particles in the droplet at the distance $r$ from the central particle in the droplet. The
green function is for  a small droplet just after the nucleation. The red curve is the distribution at the time where
the droplet has reach  a mean temperature equal to $T_{tr.p}$. The  magenta curve is $g(r)$ for a supercooled liquid at $T=0.5$ and
the blue curve is for a liquid at $T_{tr.p}$.}
\end{figure}

All 25 successful cold clusters nucleated to liquid-like droplets.
 A nucleating cluster  appears after  of the order 7 million time steps ($t_{nucl. start}= 67000_{\pm 43000}$)
and the growth to a critical nucleus  takes a mean time of $ \Delta t_{nucl}=1500_{\pm 1000 }$. 
The mean temperature during nucleation confirms the observation in \cite{toxcold}, in fact the tendency is even more
pronounced  below the triple point temperature. It is a cold small
cluster, which performs the nucleation. The nucleating cluster starts the growth with a mean temperature $T=0.47$,
and the mean temperature during the growth from a small cluster with $n_{cl} \approx 5$
particles to  the critical size with  $n_{cr.cl} \approx 18$ is  $T=0.49$, and
below the QES temperature  $T=0.50$. The  mean coordination number $n_{coor.no.}=7$ of the
particle nearest the center of mass of the nucleus corresponds to a density $\rho \approx 0.60 $, which
is equal to the density of the critical cluster at a much higher temperature \cite{toxcold}, but significantly
below the bulk density of LJ liquid ( $\rho_{tr.p}(l)$ = 0.8 and   $\rho(l) $= 0.92 (extrapolated) at $T=0.50$) \cite{Baidakov}.
 The interior of the
25 critical nuclei with $\rho \approx 0.60 $  corresponds to a diluted liquid with only a weak ordering of a first coordination shell \cite{Tox5}.
Figure 1 shows the radial distribution of particles around the central particle for a small droplet (green line) together
with the corresponding distribution at  a later time with droplet growth (red line), and the distributions
in bulk liquid at the triple point (blue line) and for supercooled liquid at $T=0.50$ (magenta line). All four curves are for a
liquid like distribution and without lattice order.

 We have also performed nucleations at other  temperatures below $T_{tr. p.}$ and for other QES densities which
confirm the result that a  nucleation below the triple point temperature
 is initiated from a small cold cluster and nucleates, as a first step  to
a liquid-like nucleus without any lattice order.

\section{Droplet growth}

The  critical nucleation is followed by  a period with liquid  droplet growth. The growth is associated with a release of latent heat which
primarily increases the temperature in the droplet, but  subsequently dissipates into the  surrounding QES gas.
An example of the  temperature evolution is shown in Figure 2. The  two curves show the fluctuating temperature in the 
droplet during the  nucleation and the succeeding growth of the liquid-like droplet.
The growth is without a thermostat (green curve) and with a Nose-Hoover thermostat (red curve). The temperature, $T_{droplet}$, of the fluid droplets
 increases dramatically and reaches $T_{droplet} \approx T_{tr.p.}$ after  $t \approx  31560$. The droplets remain, however, liquid-like during the
growth without a crystallization. The radial distribution of particles  around the central
particles in the droplet (NVE) at  $T_{droplet} \approx T_{tr.p.}$ is shown in Figure  1 (red curve).

\begin{figure}
\begin{center}
\includegraphics[width=12cm,angle=0]{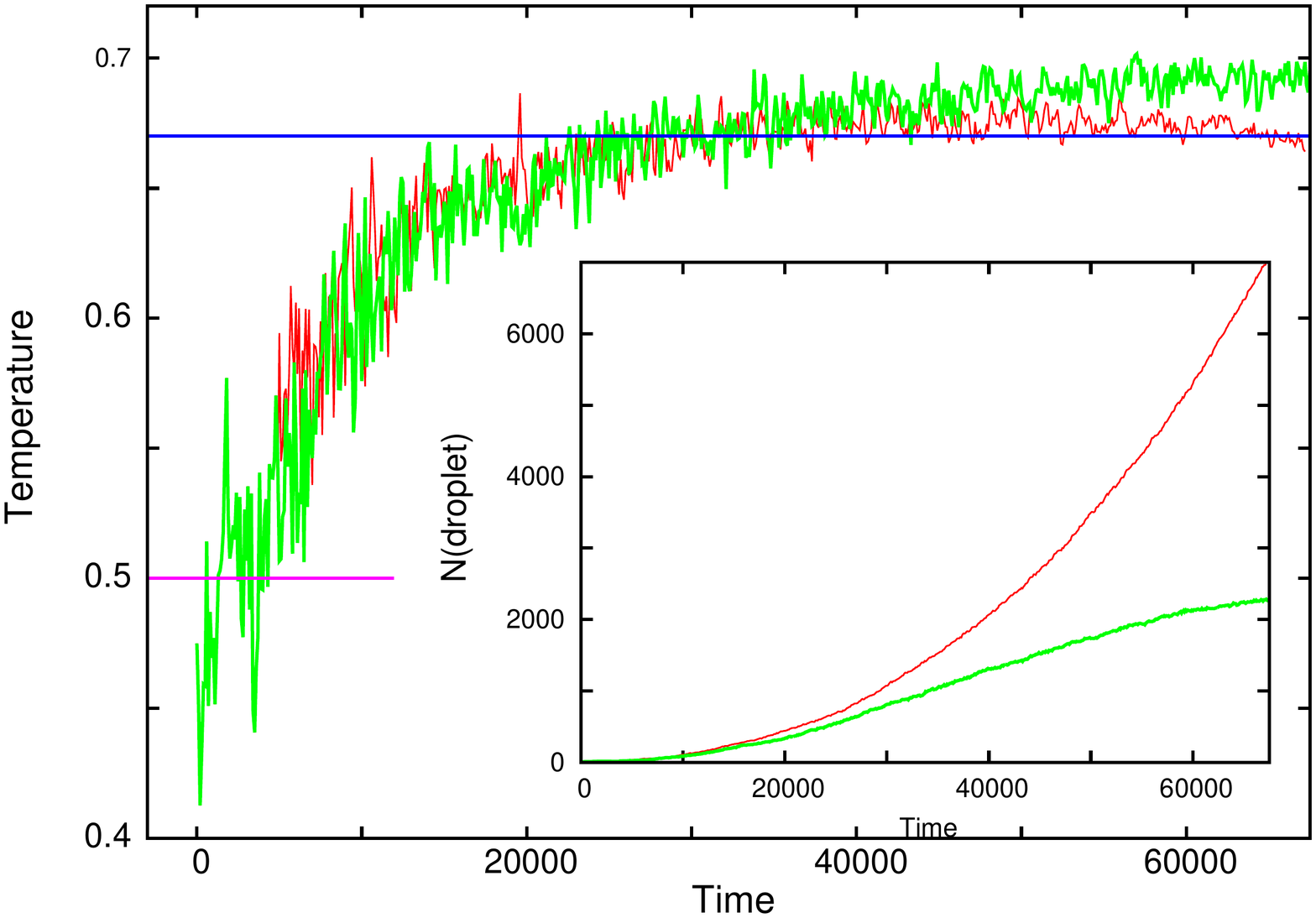}
\end{center}
\caption{Temperature evolution during nucleation and droplet growth  at $T_{\textrm{QES}}=0.50$. The green  curve shows the
temperature  in the droplet as a function of nucleation time $t$ from the
time where the sub-critical cluster starts the growth  ($t_{start}=0$). 
The red curve is  the  temperature evolution with NVT growth.
 The inset shows (green: NVE; red: NVT) the corresponding number of particles in the droplet, $n_{droplet}(t)$
 from the start and to the time  $t \approx 31560$,
where the temperature of the droplet exceeds $T_{tr. p.}=0.674$.}
\end{figure}

\begin{figure}
\begin{center}
\includegraphics[width=12cm,angle=0]{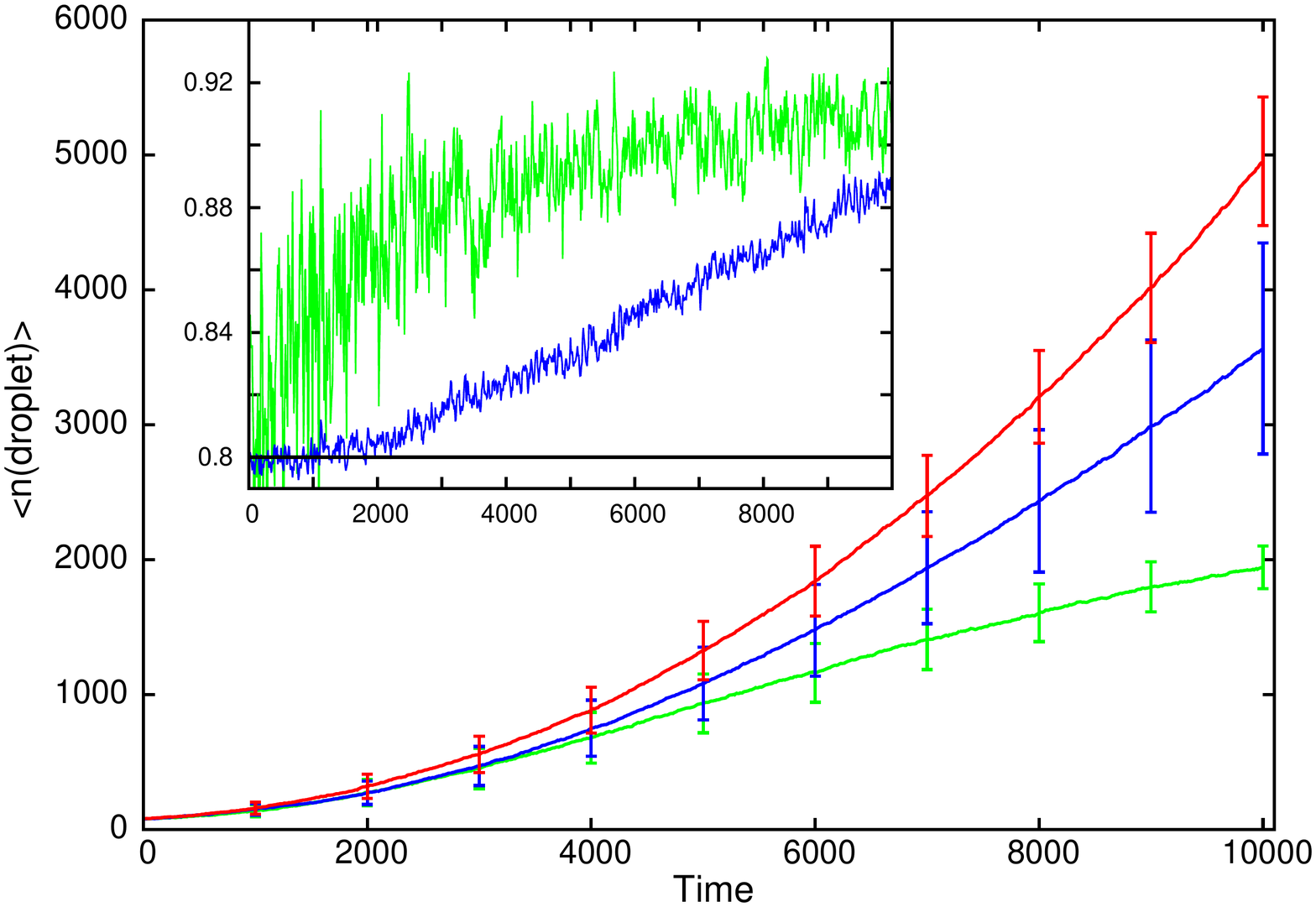}
\end{center}
\caption{ Time  evolution of the mean number, $<n_{droplet}(t)>$, of particles
 in the droplets at $T_{\textrm{QES}}=0.80$. The green curve is the mean of  25  simulations without a thermostat (NVE). The blue curve is the corresponding
mean growth, when  only the particles far away from the droplet are rescaled to maintain the temperature in  the QES gas.
 The red curve is the growth  for NVT  dynamics of the total system.
 The inset shows the corresponding temperature evolution during the NVE growth of a droplet. The green curve shows the
temperature in the droplet and the blue curve shows the temperature of
the  particles in the  gas far away from the droplet.
}
\end{figure}
 The intensive thermostat affects, however, the growth rate. The inset in Figure 2 shows the two growth and they
differ already shortly after the nucleation. The latent heat dissipates slowly into the QES gas and establishes a temperature
gradient at the droplet. The NVT simulation removes kinetic energy of a particle proportional
to its excess temperature by which  the temperature gradient is reduced. The growth is faster for a lower temperature of
the droplet,  and this fact
 explains the (nonphysical) enhanced  droplet growth by NVT simulations.

 The latent heat dissipates into the surrounding supersaturated
gas.  A  "final size" effect of the NVE simulations, due to the periodical boundaries, appears,  when the
temperature gradient exceeds half of the box length $l$. The entire system is then heated up.
 When the temperature, shown in Figure 2,   exceeds the triple point temperature,  the droplet has heated up
the gas of  particles,
 located more than $l/2$ away from the center of the droplet, from $T_{\textrm{QES}}=0.50$ to $T_{\textrm{QES}}=0.53$.

 The impact of the released  latent
heat on droplet growth is of course not limited to growth below the triple point temperature.
Figure 3 shows the corresponding temperature evolution by droplet growth at a supersaturated
gas with the temperature $T_{\textrm{QES}}=0.80$.
  The green curve with rms deviations is the mean of the twentyfive
 independent droplet growth and without a thermostat (NVE), and the red and blue curves are for different NVT dynamics. The red curve
is the growth when the velocities for all 40000 particles   are rescaled  every time step to  $T=0.80$  \cite{Diemand}, and the blue
curve is the growth, when only the velocities of the   particles outside a sphere with radius  $r_{sph}=l/2$ are rescaled.
The inset in the figure shows the temperature in a droplet (green curve) together with the
temperature $T_{\textrm{QES}}$ in the gas far away from the droplet (blue line) at the NVE simulation. The
total system begins to be heated up already after $\approx$ 200000 time steps ($t=2000$). 

The intensive thermostats reduce the local excess kinetic energy and thereby  the temperature   in the droplet and the  temperature gradient
between the droplet and the QES state.
 The  consequence of the lower excess temperature in the droplet and
 the supersaturated  gas nearby is, that the NVT growth is faster because a  particle, which collides with the droplet, has an
increased ability to be attached to the droplet.

 The system size  and the supersaturation in the present simulations are chosen in order to ensure that there appears only one
critical nucleus and droplet.
The released latent heat is proportional to the number of growing droplets, which for a given supersaturation is
proportional to the volume of the MD system, given by the number $N$ of simulated particles. So the artifact of
 an intensive thermostat is not removed by increasing the size of the system. If one e.g. increases the system size (
length of the box) by a factor of two, i.e. increases $N$ with a factor of eight and the volume with a factor of four,
 one also increases the probability of obtaining a  critical nucleus and a subsequent release of latent heat by a factor of four.

    Almost all MD  (and Monte Carlo (MC)) simulations of
droplet growth are NVT simulations with intensive thermostats which enhance the growth, and in an nonphysical way.  

\subsection{Droplet growth at less supersaturation}

The supersaturation $S=73$  for   $T_{\textrm{QES}}=0.50$ 
is  higher than the supersaturations in experiments of nucleation in  Argon \cite{Schweizer},
 and it is much higher than in a supercooled gas  in nature.
The released latent heat dissipates into the surrounding and establish a temperature gradient between the growing droplet and the  gas.
The total temperature difference between the droplet and the QES state for  $S=73$ 
 is given by the difference between the green and blue curves in the inset of Figure 3.
The droplet growth is   slower at a low supersaturation.   The consequence of this fact is,  that the released heat has a longer time to
dissipate  into the surrounding gas and with  a smaller temperature gradient between the growing droplet and the  gas far away.
It is difficult to investigate homogeneous nucleation at a low supersaturation, because the nucleation time increases exponentially
with decreasing supersaturation.
But the size of the critical nucleus  varies, however, only a little with the degree of supersaturation \cite{Diemand},
 and this fact makes it possible
to investigate droplet growth at a low supersaturation.

\begin{figure}
\begin{center}
\includegraphics[width=12cm,angle=0]{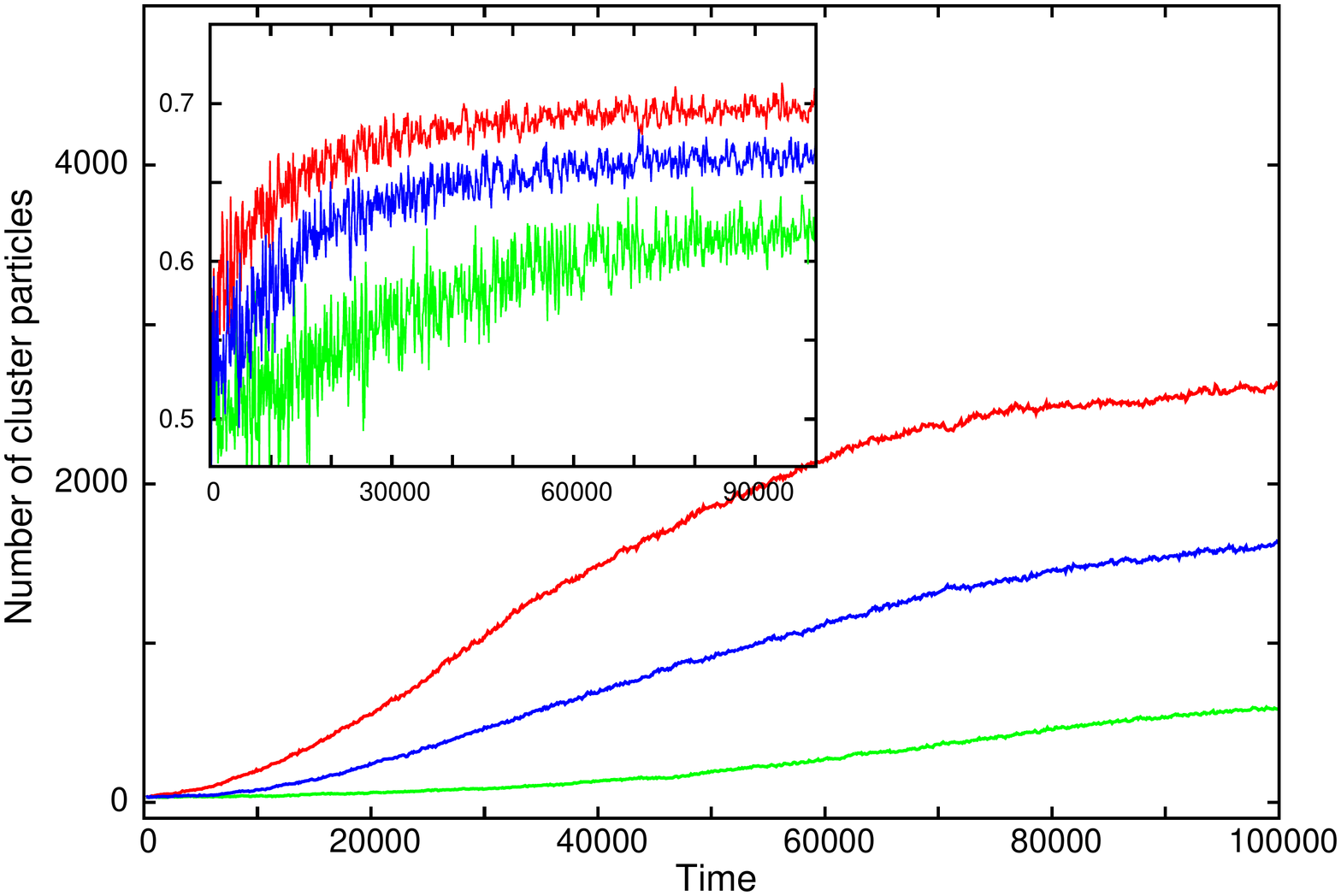}
\end{center}
\caption{Droplet growth  for different QES densities obtained by NVE dynamics. The red curve is for $\rho_{\textrm{QES}}=0.0040$; blue curve: $\rho_{\textrm{QES}}=0.0030$;
 and green curve: $\rho_{\textrm{QES}}=0.0020$. The inset shows the corresponding temperature evolutions in the droplets.}
\end{figure}

 The supersaturation is
reduced by  removing gas particle  randomly in the  gas around   the critical nucleus, once it is created.
The supersaturation is reduced to  a density $\rho=0.0030$ with $S=0.55$ and  $\rho=0.0020$ with $S=0.37$, respectively.
The droplet growth is shown in Figure 4.
The temperature in the droplet remains well below the triple point temperature for a  supersaturation $S=0.37$ (green curve), but without a crystallization
within the first  $t=10000$ time   of droplet growth, which
 corresponds to a growth time of
$\approx 1 \mu s$ for droplet growth in supersaturated Argon.
 At the time $t=10000$ the droplet with $S=37$ (green curve) contains
$n_{droplet} \approx 600 $ particles without crystallizing and despite  the fact,
 that the interior of the droplet has a bulk liquid-like mean density.
The droplet was followed for another time $t=10000$ where it continued the growth without  a crystallization.

\subsection{Crystallization of supercooled LJ droplets}

 The   vitrification and crystallization
  of nanodroplets    has been the subject of great interest for long times  \cite{Bal}.
All MD (and MC) simulations of crystallization in LJ nanodroplets have been for NVT simulations \cite{Tanaka1,Polak,Voivod,Malek}. Some of the simulated
crystallizations of nanodroplets of LJ particles have, however, been for droplets at low vapor densities and  without growth
\cite{Polak,Voivod,Malek}. The crystallization of the nanodroplets shows a complex lattice  order,
 but it is unclear, whether the  release of latent heat by crystallization might affect the crystallization. 
The debate in the literature has been, whether the crystallization is obtained as a crystallization
 in the bulk part of the droplet or is surface initiated \cite{Djikaev1,Li}.

\section{Summary}
 The thermodynamic stable state for condensed matter below the triple point temperature is a crystalline state. Never the
less the homogeneous nucleation in a supersaturated gas of LJ particles  at a temperature below the triple point temperature is to a
liquid-like critical nucleus \cite{Tanaka1,Diemand,Voivod,Malek}. The present constant energy simulations show that 25 independent nucleations
at a low temperature, $T=0.50$, well below the triple point temperature all
are initiated by a small cold cluster and nucleate to  liquid-like droplets.  The droplets remain liquid-like for a long time before
they crystallize.

 The supercooled  liquid-like glass state is one of the quasi-equilibrium states at low temperature \cite{Debenedetti} and the 
  homogeneous nucleation  to a liquid-like state  is in accordance with-, and an extension of Ostwald's step rule for phase change to a condensed  state for
a system with   polymorphism. \cite{Ostwald}.

The present MD simulations show, that the release of latent heat at the phase change affects the growth of the new phase. All simulated homogeneous  nucleations
 are with an intensive thermostat,
which suppresses the temperature gradients at the growing droplets and enhances the growth. It is, however, possible to
conserve energy in   a big MD system and for long times  by a correct treatment of the potential \cite{Tox2,Tox3,Tox4}, and 
 simulations of homogeneous nucleation and droplet growth should be without  thermostats, since nucleation and growth in nature are at constant energy.

The classical nucleation theory  for homogeneous nucleation and the succeeding droplet growth are in general formulated 
as a non-equilibrium reaction, but at a constant temperature. The present simulations reveal, that this is not correct for
the actual supersaturations considered in the MD (and MC) simulations. The  temperature
of the droplets during
the growth  will, however, deviate less from the surrounding
 gas state for less supersaturations, and the classical nucleation theory might
be considered as a limit law for homogeneous nucleation  at small supersaturations.

\section*{Acknowledgments}
The center for viscous liquid dynamics ‘Glass and Time’
is sponsored by the Danish National Research Foundation
(DNRF).


\begin{thebibliography}{99}
\bibitem{ToxMD} In MD the mass $m$ is usually included in the time unit.
 The unit length, energy and time used for LJ systems are, respectively,
 $\sigma$, $\epsilon$ and $\sigma\sqrt{m/\epsilon}$.
\end{thebibliography}
\end{document}